\documentclass[12pt]{article}
\usepackage[margin=2 cm]{geometry}
\usepackage{comment}
\usepackage{multirow}
\usepackage{graphicx}
\usepackage[font={small,it}]{caption}
\usepackage{mwe}
\usepackage[nottoc]{tocbibind}
\usepackage{amsmath,amssymb,extarrows,mathtools,graphicx,subfigure,setspace}
\usepackage{cite}
\usepackage{tikz}\usetikzlibrary{calc}
\usepackage{braket}
\usepackage{epsfig}
\usepackage[section]{placeins}
\usepackage{slashed}
\usepackage{color}
\usepackage{caption}
\usepackage{amsmath}
\usepackage{hyperref}
\makeatother

\newcommand{\be}{\begin{equation}}
\newcommand{\bea}{\begin{eqnarray}}
\newcommand{\eea}{\end{eqnarray}}
\newcommand{\ba}{\begin{array}}
\newcommand{\ea}{\end{array}}
\newcommand{\ee}{\end{equation}}
\newcommand{\bes}{\begin{equation*}}
\newcommand{\beas}{\begin{eqnarray*}}
\newcommand{\eeas}{\end{eqnarray*}}
\newcommand{\bas}{\begin{array*}}
\newcommand{\eas}{\end{array*}}
\newcommand{\ees}{\end{equation*}}

\numberwithin{equation}{section}

\begin{document}

\onehalfspacing
\vfill
\begin{titlepage}
\vspace{10mm}

\begin{center}

\vspace*{10mm}
\vspace*{1mm}
{\Large  \textbf{Modular Flow of Celestial Conformal Field Theory}} 
 \vspace*{1cm}
 
{$\text{Mahdis Ghodrati}^{a}$}

\vspace*{8mm}
{ \textsl{
$^a $ International Centre for Theoretical Physics Asia-Pacific (ICTP-AP),
University of Chinese Academy of Sciences (UCAS), Beijing, China}} 
 \vspace*{0.4cm}

\textsl{e-mails: {\href{mahdisghodrati@ucas.ac.cn}{mahdisghodrati@ucas.ac.cn}}}
 \vspace*{2mm}

\vspace*{1.7cm}

\end{center}

\begin{abstract}
We first review the calculations for the modular flow and the vector flow of $\text{CFT}_2$, warped CFTs and Bondi-Metzner-Sachs Field Theories (BMSFTs), and then we present the vector flows and modular flows in Celestial field theory and Klein CFTs. We also discuss the search for this structure in Lifshitz and other exotic field theories.

 \end{abstract}

\end{titlepage}

\tableofcontents

\section{Introduction}
It is already an established fact in holography that quantum entanglement can reconstruct the bulk geometry from boundary data, and it is a general principle beyond AdS/CFT. Based on various studies such as those in \cite{Apolo:2020bld, Apolo:2020qjm}, the entanglement entropy satisfies some consistency conditions \cite{Czech:2025jnw}, and these conditions can lead to the reconstruction of kinematic space. Then, kinematic space reconstructs the bulk metric, and the bulk Einstein equations follow from entanglement. In any of these steps, conformal symmetry or Lorentz invariance is not necessary and only considering entanglement associated with a boundary region would be enough. So, the notions of generalized Ryu-Takayanagi surfaces, Wilson lines, flat-space geodesics, and modular Hamiltonians could be defined for WCFTs and BMSFTs. Note that these two field-theory models have an infinite set of local symmetries. This has been done in \cite{Apolo:2020qjm}.

The modular Hamiltonian, which is a non-local operator, is an important tool in this study which is defined as
\begin{gather}
\mathcal{H}_{mod} = - \log \rho_{\mathcal{A}},
\end{gather}
where $\rho_{\mathcal{A}}$ is the reduced density matrix of the state on a subregion $\mathcal{A}$.

For the study of symmetry transformations, the modular Hamiltonian is also important as it generates a transformation between operators in the causal domain of $\mathcal{A}$.  Also, it has been shown that the diffeomorphism generated by modular flow generator $\zeta$ maps each point in the causal domain $\mathcal{D}$ back to itself.  Here, $\zeta$ leaves the causal domain $\mathcal{D}$ of the interval invariant, and the flow is given by $e^{i\zeta}$.

Because of these interesting properties of the modular Hamiltonian regarding its connections to symmetries and bulk reconstruction from boundary data, it would be interesting to generalize it to other interesting field theories with exotic properties such as those without Lorentz symmetry (e.g., Lifshitz) , as well as new proposed models like celestial CFTs and Klein CFTs.

Here we aim to first construct the modular flow and vector flow for the case of CCFTs, where the boundary theory lives on the celestial sphere, $S^2$, or $\mathbb{CP}^1$, instead of a timelike boundary, and the bulk is asymptotically flat. For this case, instead of AdS geodesics, one could consider null sheets, or light-ray operators, or causal diamonds, and then build null kinematic space and subsequently the bulk.

Another interesting example is the Klein CFT, which have non-orientable worldsheets, crosscaps, involutions, and orientation reversal. Their holographic duals are expected to involve quotient geometries, orbifolds, non-orientable bulk manifolds, and crosscap states.

The modular Hamiltonian and modular flow could also be used for celestial CFTs and Klein CFTs to understand bulk reconstruction from boundary data. So, the kinematic space, which has been used for the bulk reconstruction can be replaced by null kinematic space for the case of celestial CFTs and quotient kinematic space for Klein CFTs. However, one should note that we still do not have a universally accepted analogue of the Ryu-Takayanagi prescription for Klein CFTs or CCFTs that connect extremal surfaces in non-orientable bulk space to entanglement in the boundary. So for the case of CCFTs, null extremal surfaces or light sheets should be constructed, and for the case of Klein holography, quotient, or crosscap-sensitive extremal surfaces should be considered. For this aim, fully understanding the modular structures in these theories is important, which is the aim of this work.

\section{The setup}

Considering a region $A$ and its complement $\bar{A}$, with the whole Hilbert space given by $\mathcal{H} = \mathcal{H}_A \otimes \mathcal{H}_{\bar{A}}$, the reduced density matrix can be written as $\rho_A \equiv \text{Tr}_{\mathcal{H}_{\bar{A}} } \rho $ where $\rho$ is the density matrix of the entire system, and the entanglement entropy of this pure state would be $S_A \equiv - \text{Tr}_{\mathcal{H}_A} ( \rho_A \log \rho_A)$. 

Using the terminology of modular Hamiltonian, denoted by $K_A$, the reduced density matrix is $\rho_A = e^{- K_A} / \mathcal{Z}_A$ where $\mathcal{Z}_A = \text{Tr}_{\mathcal{H}_A} (e^{-K_A}) $ is a constant leading to the normalization condition $\text{Tr}_{\mathcal{H}_A } \rho_A=1$.

For a sphere $B$ with radius $\mathcal{R}$ in $d+1$-dimensional CFT, in Minkowski spacetime and in its ground state, the modular Hamiltonian is
\begin{gather}
K_B= \mathcal{R} \int_B \beta(r) \ T_{tt}( \boldsymbol{x}) d^d \boldsymbol{x},
\end{gather}
where the weight function $\beta(r)$ is a parabola of the form
\begin{gather}
\beta(r) \equiv 2\pi \left \lbrack \frac{1}{2} \left ( 1- \frac{r^2}{ \mathcal{R}^2} \right ) \right \rbrack.
\end{gather}

Now the aim is to find the form of this weight function for warped CFTs, BMS, celestial field theory, and some other exotic field theories and check how the algebra and specific properties of such field theories affect the form of these weight functions and the whole modular flow structure.

Here, we show the modular flow and vector flow in celestial field theory and Klein CFTs, and compare them with the flow structure of CFTs, WCFTs and BMSFTs.

\section{Modular Hamiltonian and modular flow for exotic field theories}
Here, we first review the calculations of modular Hamiltonian and modular flow in common field theory cases such as $\text{CFT}_2$, BMSFTs, and WCFTs, and we present the modular flow structure in some other exotic examples of field theories, including the case of CCFT, Klein CFT, and we discuss the Lifshitz case.

\subsection{Modular flow of $\text{CFT}_2$} 

The analysis of modular flow here is for the ``two-sided'' vacuum modular Hamiltonian of an interval. The modular Hamiltonian of two-sided vacuum $\text{CFT}_2$ case has global $SO(2,2)$ symmetry algebra, which for a strip is decomposed to a pair of commuting $SO(2,1)$ subalgebras, acting on the left- and right-moving null coordinates $x^+$ and $x^-$, with the commutation relations given by
\begin{gather}
\lbrack L_0, L_1 \rbrack = - L_1 \ \ \ \ \ \ \lbrack L_0, L_{-1} \rbrack= L_{-1} \ \ \ \ \ \ \ \lbrack L_1, L_{-1} \rbrack = 2L_0,
\end{gather}
and similarly for $\bar{L}_i$.

The modular Hamiltonian for an interval with endpoints $x_L^\mu = (a^+, a^-)$ and $x_R^\mu=(b^+, b^-)$ is a boost that preserves $x_L$ and $x_R$.  It has the form $H_{\text{mod}}=K_+ + K_- $, where $K_+$ and $K_-$ are linear combinations of $L_i$s and $\bar{L}_i$s. These operators can be written as
\begin{gather}
K_+= s_1 L_1 + s_0 L_0+s_{-1} L_{-1}, \nonumber\\
K_-= t_1 \bar{L}_1 + t_0 \bar{L}_0+t_{-1} \bar{L}_{-1}.
\end{gather}

So these coefficients in the CFT case can be determined up to an overall multiplicative constant, since the only requirement is that the generators $K_+$ and $K_-$ preserve the left- and right-moving null coordinates of the interval endpoints. 
Equivalently, the requirement is that the operator $\exp (-H_{\text{mod} }/2)$, which is a finite $SO(2,1) \times SO (2,1)$ transformation, should map an interval to its complement.

In \cite{Apolo:2020qjm}, for the WCFT case, this condition has been written as that the modular flow generator $\zeta$ vanishing along the boundaries  $\partial \mathcal {D}$ of the causal domain, or as $\zeta \big |_{\partial \mathcal{A}} \propto \partial_\omega $, which essentially is the same as the above differential equations. In \cite{deBoer:2021zlm}, the parallel transport of the modular Hamiltonian, which encodes the entanglement properties of states, has been studied.

If one writes the generators in the following representation \cite{Czech:2019vih},
\begin{gather}
L_{-1}=i e^{-i x^+} \partial_+, \quad L_0=i \partial_+, \quad  L_1=i e^{ix^+} \partial_+,
\end{gather}
then, using the condition mentioned, one can then find \cite{Czech:2019vih},
\begin{gather}
s_1= \frac{2\pi \cot ( (b^+-a^+)/2)}{e^{ia^+}+e^{i b^+}}, \quad  s_0=-2 \pi \cot(b^+-a^+)/2, \quad s_{-1}=\frac{2\pi \cot(b^+-a^+)/2}{e^{-ia^+}+e^{-ib^+} }, \nonumber\\
t_1=- \frac{2\pi \cot (b^- - a^-)/2}{e^{i a^-} + e^{i b^-} },\quad   t_0=2\pi \cot(b^- - a^-)/2, \quad  t_{-1}=- \frac{2\pi \cot(b^- - a^-)/2 }{e^{-i a^-} + e^{-i b^-} },
\end{gather}
which determines $H_{\text{mod}}$ up to an overall multiplicative constant.

\subsection{Modular flow of BMSFTs} 

The modular Hamiltonian and modular Berry curvature and connection for the case of BMS field theory (BMSFT) and warped conformal field theories (WCFTs) have been studied in \cite{Apolo:2020qjm}

For BMSFTs, the interval with endpoints can be written as \cite{Apolo:2020qjm},
\begin{gather}
\partial \mathcal{A}= \{ (u_-, z_-), (u_+, z_+) \}, \quad l_u\equiv u_+ - u_-, \quad  l_z \equiv z_+ - z_-,
\end{gather}
where $(u,z)$ are the coordinates on the plane.

First, the modular flow generator can be written as \cite{Apolo:2020qjm},
  
 \begin{gather}\label{eq:generatorform}
 \zeta= \zeta^\mu \partial_u +\zeta^z \partial_z= \sum_{j=-1}^1 ( a_j \ell_j+b_j m_j), 
 \end{gather}
where the vacuum symmetry generators on the plane for $j \in \{-1,0,1 \}$ are
\begin{gather}
\ell_j= - z^{j+1} \partial_z - (j+1) z^j u \partial_u, \quad m_j=- z^{j+1} \partial_u,
\end{gather}
with the BMS algebra
\begin{gather}
\lbrack \ell_j, \ell_k \rbrack= (j-k) \ell_{j+k}, \quad  \lbrack \ell_j, m_k \rbrack = (j-k) m_{j+k}, \quad \lbrack m_j, m_k \rbrack =0.
\end{gather}

These coefficients $a_j$ and $b_j$ can be found from the requirement that the operator $e^{i \zeta}$ maps any point in the domain of the interval $\mathcal{D}$ back to itself, or that the boundary of the causal domain $\partial \mathcal{D}$ is invariant under the modular flow.

Again, the requirement is that under the modular flow, the boundary $\partial \mathcal{D}$ is invariant. So, by imposing the condition that the modular flow generator vanishes at the endpoints, one gets
\begin{equation}
\begin{aligned}
(a_+, a_0, a_-) &= a_+ ( 1, - z_- - z_+, z_+ z_-),\nonumber\\
(b_+, b_0, b_-) & = b_+ (1, -z_- - z_+, z_+ z_-) + a_+ ( 0, - u_{-} - u_{+} , u_+ z_- + u_- z_+ ).
\end{aligned}
\end{equation}

Then, one should note that under the modular flow $e^{s\zeta}$, $z(s)$ and $u(s)$ obey the differential equations as
\begin{gather}
\partial_s u(s)= \zeta^u, \quad  \partial_sz(s)=\zeta^z,
\end{gather}
where $\zeta^u$ and $\zeta^z$ can be found from the relation \ref{eq:generatorform}.

The solution for $z(s)$ is
\begin{gather}
z(s)=\frac{e^{a_+ z_+ s + c_0 z_-  }z_+ - e^{a_+ z_- s + c_0 z_+ }z_-  }{ e^{a_+ z_+ s + c_0 z_-}-e^{a_+ z_- s + c_0 z_+}  },
\end{gather}
Here $c_0$ is a constant.

The solution for $u(s)$ in \cite{Apolo:2020qjm}, is
\begin{gather}
u(s)= \frac{e^{a_+ z_+ s}u_+ - e^{a_+ z_- s} u_- }{e^{a_+ z_+ s}-e^{a_+ z_- s} }+ \frac{2\pi l_u/l_z^2 \ . \ e^{a_+ (z_+ +z_-)s/2 } \sinh (a_+(z_+ -z_-)s/2) }{(e^{a_+ z_+ s} -e^{a_+ z_- s} )^2}  \cdot  \frac{(z_+ -z_-)^2}{2},
\end{gather}

 Then, one should apply the requirement $z(s)=z(s+i)$ and $u(s) = u(s+i)$. Using these, one can find $a_+$ and $b_+$ up to an overall sign as
 \begin{gather}
 a_+ = \frac{2\pi}{z_+ - z_- }, \quad b_+= - \frac{2\pi (u_+ - u_- )}{(z_+ - z_- )^2}.
 \end{gather}

So the coefficients are completely determined as
\begin{equation}
\begin{aligned}
(a_+, a_0, a_-)&=\frac{2\pi}{z_+ - z_-} (1, - z_- - z_+, z_+ z_-),\nonumber\\
(b_+, b_0, b_-)&= \frac{2\pi}{ (z_+ - z_-)^2} ( u_- - u_+, 2 u_+ z_- - 2 u_- z_+, u_- z_+^2 - u_+ z_-^2).
\end{aligned}
\end{equation}

Then, the modular flow generator $\zeta$ for the interval is
\begin{gather}
\zeta = \sum_{j=-1}^1 (a_j \ell_j+b_j m_j)= \lbrack T(z)+u Y^\prime(z) \rbrack \partial_u + Y(z) \partial_z,
\end{gather}
where 
\begin{equation}
\begin{aligned}
\text{for}& \ \   z\in \lbrack z_-, z_+ \rbrack \to  \ \ \ \ \ T(z)=\frac{2\pi \lbrack  u_+ (z-z_-)^2 - u_- (z- z_+)^2  \rbrack} {(z_+ - z_-)^2 }, \nonumber\\ &
 \text{otherwise} \to  \ 0,
\end{aligned}
\end{equation}
and
\begin{equation}
\begin{aligned}
\text{for}& \ \   z\in \lbrack z_-, z_+ \rbrack \to  \ \ \ \ \ Y(z)=-\frac{2\pi (z-z_-) (z- z_+) } {z_+ - z_- }, \nonumber\\ &
 \text{otherwise} \to  \ 0.
\end{aligned}
\end{equation}

\begin{figure}[ht!]   
\begin{center}
\includegraphics[width=0.4\textwidth]{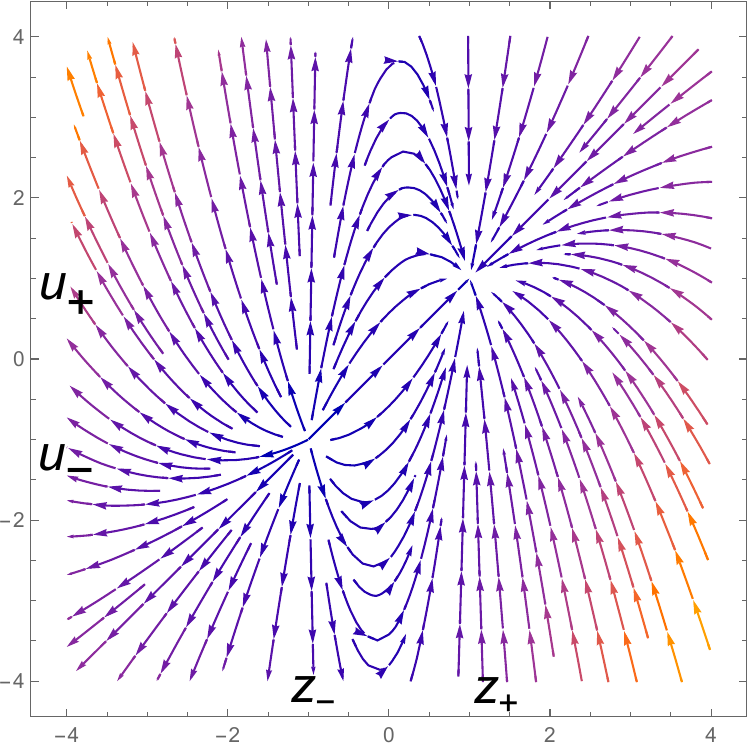}  \ \ \ \ \ \ \  \ \  \ \ \ 
\includegraphics[width=0.4\textwidth]{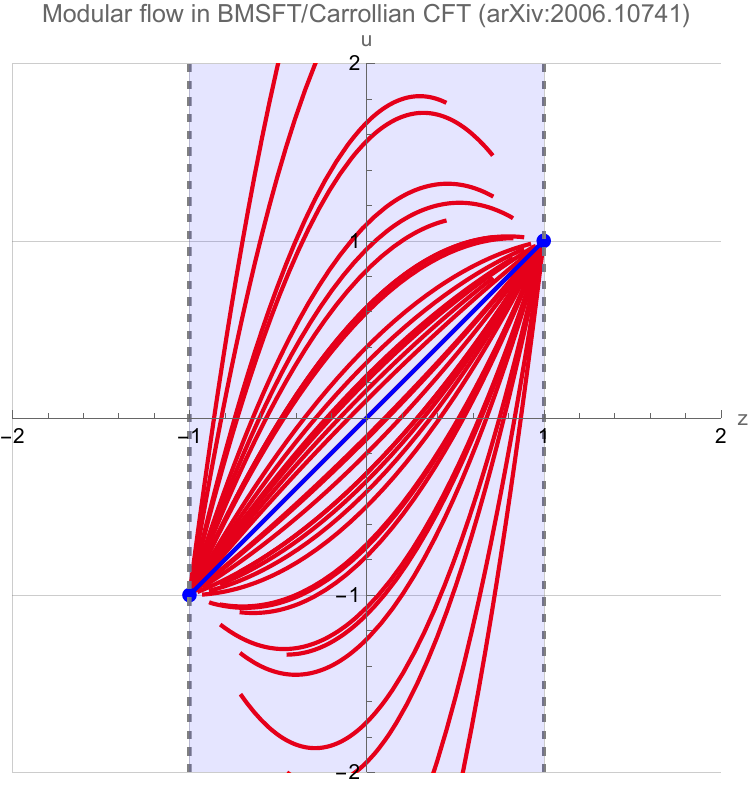}
\caption{The modular flow of BMSFT for an interval between $z_-$ and $z_+$.}
\label{fig:BMSFTModular}
\end{center}
\end{figure}

The modular flow of this field theory is also depicted in Figure \ref{fig:BMSFTModular} as well as in \cite{Apolo:2020qjm}.

\subsection{Modular flow of WCFTs} 

In \cite{Apolo:2020qjm}, the modular flow generators for warped CFTs (WCFTs) on the plane and the (thermal) cylinder have been calculated and depicted. The modular flow generator for WCFTs can be geometrically described as the combination of the $SL(2,R) \times U(1)$ generators that leave the vacuum invariant.

\subsubsection{Modular flow of WCFTs on the plane}

The interval could be written as
\begin{gather}
\partial \mathcal{A} = \{ (z_-, w_-), (z_+, w_+) \}, \ \ \ \ \ l_z\equiv z_+ - z_-,  \ \ \ l_w \equiv w_+ - w_-,
\end{gather}
and again one has the relation $(z, w) \sim (z, w-2\pi i \mu)$.

Based on the symmetries of the theory, the modular generator is a linear combination of global $SL(2,R) \times U(1)$ generators that maps causal domain of $\mathcal{A}$ into itself.

On the plane, these global symmetry generators can be written as
\begin{gather}\label{eq:wcftgenerators}
\ell_1= -z^2 \partial_z, \ \ \ \ \ \ \ell_0 = - z \partial_z, \ \ \ \ \ \ell_{-1} = - \partial_z, \ \ \ \ \ \bar{\ell}_0= - \partial_w,
\end{gather}
and the modular flow generator $\zeta$ is
\begin{gather}
\zeta = \sum_{i=-1}^1 a_i \ell_i + \bar{a}_0 \bar{\ell}_0.
\end{gather}

The coefficients $a_i$ can again be found using the condition that $\zeta$ vanishes along the boundaries $\partial \mathcal{D}$, or as $\zeta \big |_{\partial \mathcal{A}} \propto \partial_w$ leading to
\begin{gather}
a_1 =  c_0 \frac{1}{z_+ - z_-}, \ \ \ \ \ \ \ a_0= - c_0 \frac{z_+ + z_-}{z_+ - z_-}, \ \ \ \ \ \ a_{-1}=c_0 \frac{z_+ z_-}{z_+ - z_-},
\end{gather}
where again $c_0$ is a constant. This constant can be found from the requirement that $e^{i\zeta/2}$ maps a point in the domain $\mathcal{D}$ to its complement and back to itself. Considering a point $( z(0), w(0))= (\frac{z_- e^{d_0 z_+}- z_+ e^{d_0 z_-}   }{e^{d_0 z_+}-e^{d_0 z_-} }, w)$, where $d_0$ is a real constant, and since under modular flow we can write the following equality
\begin{gather}
(z(s), w(s)) = \left ( \frac{z_- e^{d_0 z_+} - z_+ e^{c_0 s} e^{d_0 z_-} }   {e^{d_0 z_+} -e^{c_0 s} e^{d_0 z_-} }, w- \bar{a}_0 s \right ),
\end{gather}
then, by using $z(0)=z(i)$, we find $c_0= 2 \pi$, which leads to
\begin{gather}
a_1= 2\pi \frac{1}{z_+ - z_-}, \ \ \ \ \ \ \ a_0= - 2\pi \frac{z_+ + z_-}{z_+ - z_-}, \ \ \ \ \ \ a_{-1}=2\pi \frac{z_+ z_-}{z_+ - z_-}.
\end{gather}

The parameter $\bar{a}_0$ then can be found by the replica symmetry in WCFTs as $(z, w) \sim (z e^{2\pi i}, w-2\pi i \mu)$, which means that when a point in $\mathcal{D}$ under the modular flow is mapped to itself, the coordinate $w$ must also be translated by the amount $2 \pi \mu$. This leads to
\begin{gather}
\bar{a}_0= 2 \pi \mu.
\end{gather}

Then, inserting the $SL(2,\mathbb{R}) \times U(1)$ generators \ref{eq:wcftgenerators}, the generator of the modular flow can be written as
\begin{gather}
\zeta= 2 \pi \mu \bar{\ell}_0 + \frac{2\pi}{z_+ - z_-} \lbrack \ell_1 - (z_+ + z_- ) \ell_0 + z_+ z_- \ell_{-1} \rbrack \nonumber\\
= - 2\pi \mu \partial_w - \frac{2\pi}{z_+ - z_-} \lbrack z_+ z_- - (z_+ + z_- ) z + z^2 \rbrack \partial_z. 
\end{gather}

\begin{figure}[ht!]   
\begin{center}
\includegraphics[width=0.4\textwidth]{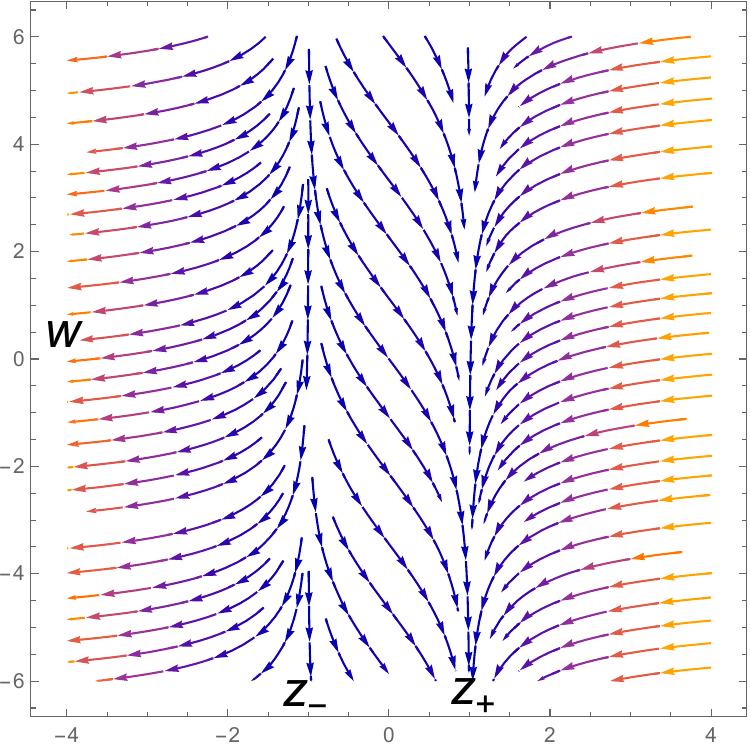}  \ \ \ 
\caption{The modular flow of WCFT for an interval between $z_-$ and $z_+$.}
\label{fig:WCFTflow}
\end{center}
\end{figure}

This is the modular flow on the reference plane. Its behavior is shown in Figure \ref{fig:WCFTflow}.

The modular flow on the thermal cylinder can also be calculated.

\subsubsection{Modular flow of WCFTs on thermal cylinder}
For the thermal cylinder, one has the identification
\begin{gather}
(x,y) \sim (x+ i \beta, y- i \theta), 
\end{gather}
so, using the warped conformal transformation \cite{Song:2016gtd}
\begin{gather}
z= e^{\frac{2\pi x}{\beta} }, \ \ \ \ \ \ \ \ w=y+ \left ( \frac{\theta-2\pi \mu }{\beta} \right)x, 
\end{gather}
one obtains the modular flow generator on the thermal cylinder as
\begin{gather}
\zeta= - 2 \pi \mu \partial_y - \beta \Bigg \lbrack  \frac{\cosh \frac{\pi (2x-x_+ - x_-) }{\beta}- \cosh \frac{\pi l_x}{\beta}  }{\sinh \frac{\pi l_x}{\beta} } \Bigg \rbrack \Big ( \partial_x - \frac{\theta-2\pi \mu}{\beta} \partial_y \Big ),
\end{gather}

Note that the phase transitions of $\text{WAdS}_3/\text{WCFT}_2$ have been studied in \cite{Ghodrati:2016ggy,Ghodrati:2022lnd}, in different massive gravity models, and the conserved charges have been studied in \cite{Ghodrati:2016vvf}. The complexity growth rate has also been studied in \cite{Ghodrati:2019bzz}.

\subsection{Modular Hamiltonian and modular flow for celestial conformal field theories}

The goal of celestial holography is to establish a holographic duality between quantum gravity in four dimensional asymptotically flat spacetime and a two-dimensional conformal field theory, i.e., celestial CFT, that lives on the celestial Riemann surface at the boundary.

In fact, the celestial holography dictionary connects scattering particle in the bulk spacetime to operator $\mathcal{O}_{\Delta, J}$ in CCFT that live null infinity.

Now, using a similar procedure, we aim to derive the modular flow generator and modular Hamiltonian for celestial field theories \cite{Raclariu:2021zjz,Donnay:2021wrk, Donnay:2020guq}, Klein CFTs, and discuss the Lifshitz case.

Note that the modular flow in celestial conformal field theory is associated with the entanglement structure of quantum fields living on the celestial sphere, which is the boundary of $4D$ asymptotically flat spacetime. 

The celestial theory inherits most of the structure of the extended $\text{BMS}_4$ algebra. These structures include a global conformal group $SL(2, \mathbb{C})$ that acts on the celestial coordinates $(z, \bar{z})$, and the supertranslations that are angle-dependent translations at null infinity. 

Note that Carrollian CFT and flat holography can be related to celestial holography by integrating out the null direction \cite{Donnay:2022wvx, Donnay:2022aba, Bagchi:2022emh,Pasterski:2017kqt}. So, by first by considering modular flow in Carrollian CFT, we can move to modular flow in celestial conformal field theory.

The global symmetry generators of Carrollian CFT on the plane are shown in the following table. 

\begin{center}
\begin{tabular}{||c c c ||} 
 \hline
 Generator & Expression & Interpretation  \\ [0.5ex] 
 \hline\hline
$L_{-1} ,L_0, L_1$ & $-\partial_z, - z \partial_z, -z^2 \partial_z$ & 	Holomorphic global conformal transformations  \\
 \hline
$\bar{L}_{-1} , \bar{L}_0, \bar{L}_1$ & $-\partial_{\bar{z}}, - \bar{z} \partial_{\bar{z}}, -\bar{z}^2 \partial_{\bar{z}}$ &Antiholomorphic global conformal transformations  \\
 \hline
$T_1, T_z, T_{\bar{z}}, T_{z \bar{z}}$ & $\partial_u, z \partial_u, \bar{z} \partial_u, z \bar{z} \partial_u$ & Global $4d$ translations (supertranslations)
  \\
 \hline
\end{tabular}
\end{center}

For celestial field theory, the time coordinate should then be omitted. 

For a single interval on the celestial plane (stereographic projection of the sphere) with complex coordinate $(z, \bar{z})$ and the retarded time at future null infinity $J^+$ denoted by $u$, where the field theory lives at fixed large radius $r$, one can write the modular flow generator.

For the interval $l \subset (z, \bar{z}) $ between two points $(z_1, \bar{z_1} )$ and $(z_2, \bar{z_2} )$ at fixed $u$, the general form of the modular flow generator vector field is
\begin{gather}
\zeta = \xi^z (z) \partial_z + \xi^{\bar{z}} (\bar{z}) \partial_{\bar{z}} + \xi^u (z, \bar{z}, u) \partial_u,
\end{gather}
where the $\xi^z$ and $\xi^{\bar{z}}$ are the holomorphic and antiholomorphic parts.

Since the conformal part of modular flow preserves the interval points, similar to the case of $2D$ CFT, we have
\begin{gather}
\xi^z (z) = 2\pi  \frac{(z(s)-z_1) ( z_2 -z(s))}{z_2-z_1},\nonumber\\
\xi^{\bar{z}} (\bar{z}) = 2\pi  \frac{(\bar{z} (s) -\bar{z}_1) ( \bar{z}_2 -\bar{z}(s))}{\bar{z}_2-\bar{z}_1},
\end{gather}
which generate $SL(2, \mathbb{C})$ boosts across the interval.

Then, for the supertranslation part or the null time $\xi^u$, we should consider that in celestial conformal field theory, $u$ transforms under Lorentz transformations $(SL(2, \mathbb{C}))$ as well as under supertranslations, and under infinitesimal conformal transformations on the celestial sphere, the shift of $u$ is $u \to u+ f(z, \bar{z})$.

In order to preserve the causal structure of the interval, the modular flow should include a $u$-dependent component in the form of 
\begin{gather}
\xi^u (z, \bar{z}, u) = 2\pi (u - u_0) v(z, \bar{z}).
\end{gather}

The function $v(z, \bar{z})$ vanishes at the endpoints and $u_0$ is a fixed reference time. A natural choice for $v(z, \bar{z})$ is
\begin{gather}
v(z, \bar{z}) = \frac{(z-z_1) (z_2 - z) }{z_2 - z_1} . \frac{(\bar{z}-\bar{z}_1) (\bar{z}_2 - \bar{z}) }{\bar{z}_2 - \bar{z}_1}, 
\end{gather}

Therefore, one gets
\begin{gather}
\xi^u ( z, \bar{z}, u) = 2\pi ( u-u_0) \frac{(z-z_1) (z_2-z)}{z_2 - z_1} . \frac{(\bar{z}-\bar{z}_1) (\bar{z}_2 - \bar{z}) }{\bar{z}_2 - \bar{z}_1}.
\end{gather}

The final form of the modular flow generator in Carrollian CFT is
\begin{gather}
\zeta = 2\pi \frac{ (z-z_1) ( z_2 - z)}{z_2 - z_1} \partial_z+ 2\pi \frac{(\bar{z} - \bar{z}_1) (\bar{z}_2 - \bar{z} ) }{\bar{z}_2 - \bar{z}_1} \partial_{\bar{z}} +   
2 \pi ( u-u_0) \frac{(z-z_1)(z_2-z) }{z_2 - z_1}   \frac{(\bar{z}-\bar{z}_1)(\bar{z}_2-\bar{z}) }{\bar{z}_2 - \bar{z}_1}\partial_u, 
\end{gather}
and for celestial CFT, it is
\begin{gather}
\zeta = 2\pi \frac{ (z-z_1) ( z_2 - z)}{z_2 - z_1} \partial_z+ 2\pi \frac{(\bar{z} - \bar{z}_1) (\bar{z}_2 - \bar{z} ) }{\bar{z}_2 - \bar{z}_1} \partial_{\bar{z}}. 
\end{gather}
Now we can plot these two and compare them. Note that celestial CFT lives on a celestial sphere and Carrollian CFT lives on null infinity.

\begin{figure}[ht!]   
\begin{center}
\includegraphics[width=0.4\textwidth]{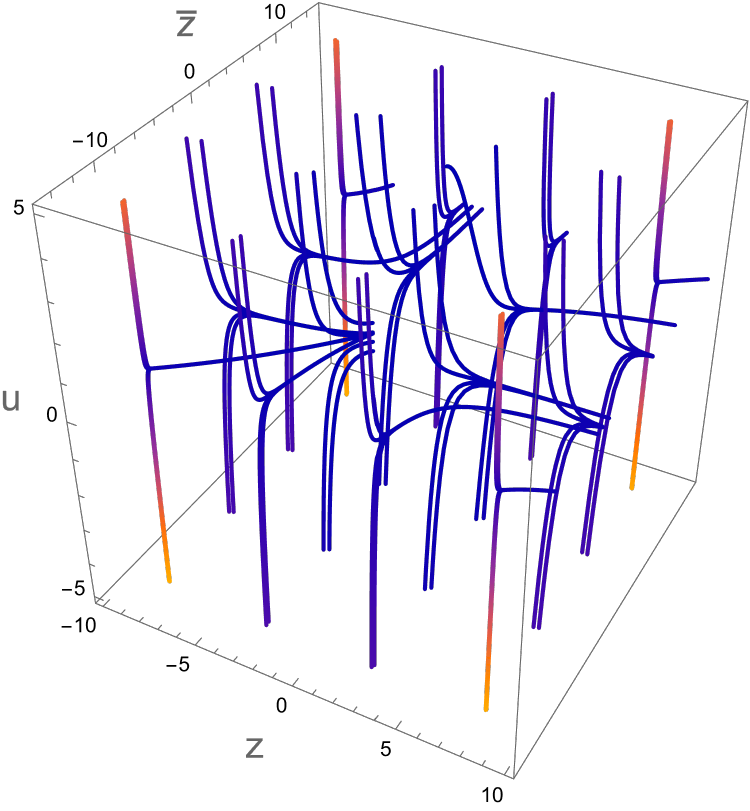}  \ \ \ 
\includegraphics[width=0.4\textwidth]{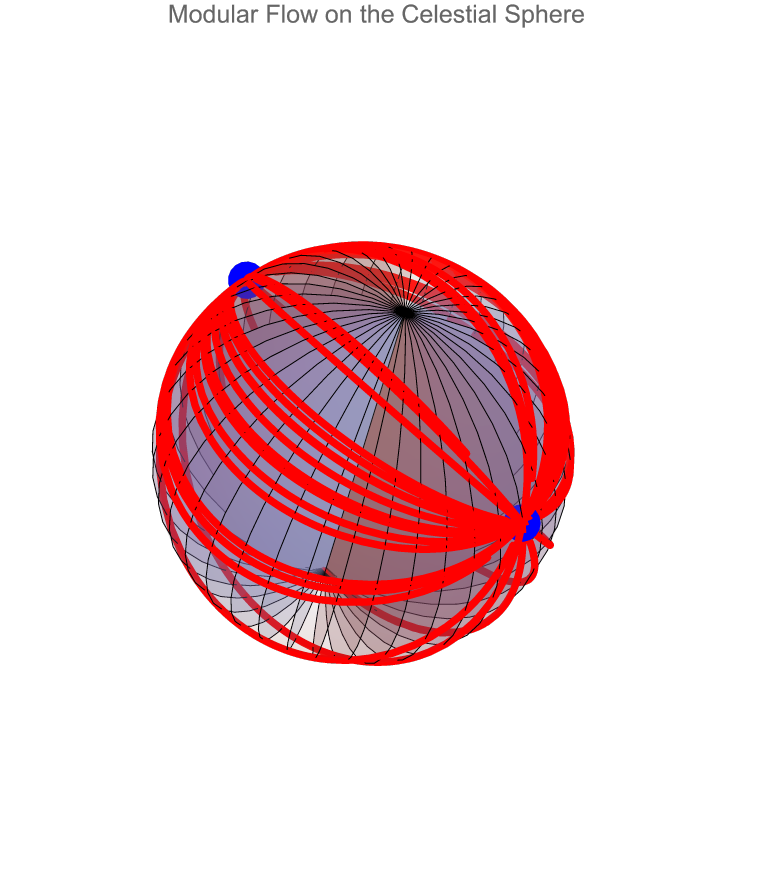}  
\caption{On the left: the modular flow of celestial CFT for an interval between $\mathcal{J} \subset (z, \bar{z})$, which is between two points $(z_1, \bar{z_1})$ and $(z_2, \bar{z_2}) $ and at fixed $u$. On the right: the modular flow on the celestial sphere.}
\label{fig:CCFTmodularf}
\end{center}
\end{figure}

The vector structure of modular flow of celestial conformal field theory is shown in Figure \ref{fig:CCFTmodularf}.

The form of modular Hamiltonian is
\begin{gather}
K= \int_{\text{interval}} dz d\bar{z} \left ( \xi^z(z) T(z) + \xi^{\bar{z}} (\bar{z}) T(\bar{z}) + \xi^u ( z, \bar{z}, u) M(z, \bar{z} )      \right)
\end{gather}
where $T(z)$ and $T(\bar{z})$ are the components of $2D$ celestial stress-energy tensor, and $M(z, \bar{z})$ is the operator generating supertranslations.

Note that $\partial_z$ and $\partial_{\bar{z}}$ generate local $SL(2,\mathbb{C})$ boosts across the interval, and $\partial u$ is the supertranslations part that respects the causal structure along $u$ (null direction). In addition, the modular flow preserves the causal diamond of the interval on $J^+$.

\subsection{Klein CFTs}

The conformal symmetries of celestial conformal field theories are induced by superrotations which are part of extended Bondi-Metzner-Sachs (BMS) asymptotic symmetries in the bulk theory. In this duality, the dictionary is that the Ward identities of the gravitational S-matrix for the extended BMS symmetries that correspond to soft theorems in the bulk are dual to $2D$ currents in celestial conformal field theory.

This can be considered as the analytic continuation from Minkowski space to $(2,2)$ split-signature spacetime, where null infinity becomes the product of null interval with a celestial torus \cite{Atanasov:2021oyu}. Also, note that here $L_0 \pm \bar{L}_0$ generate the compact space and time directions of the celestial torus and are quantized. These groups preserve the $\text{AdS}_3/ \mathbb{Z}$ hypersurfaces, which are a fixed distance from the origin and foliate the Klein space. 

Note that in \cite{Atanasov:2021oyu}, some modes corresponding to particles emerge at a fixed point on the celestial torus, where they correspond to the H-primary operators, and the scattering of such particles takes the form of a correlation function on the celestial torus. Now the modular flow in this theory also follows such trajectories. The H-primary wave functions which have quantized weights, form a torus $\mathbb{C}/(\mathbb{Z} + \tau \mathbb{Z})$ and the L-primary wave functions can be expressed as weighted integrals over such torus. So modular flow can be written as the spectral flow in the H-primary expansion.

Here we use the generators given in \cite{Atanasov:2021oyu} for the Lorentz group of $\mathbb{K}^{2,2}$ which is $SO(2,2) \sim \frac{SL(2, \mathbb{R})_L \times SL(2, \mathbb{R})_R}{\mathbb{Z}_2} $, where the $\mathbb{Z}_2$ is generated by $-1_L \times -1_R$, and the spin group is the double cover $SL(2, \mathbb{R})_L \times SL(2, \mathbb{R})_R$.

The symmetries of this theory are generated by the real combinations of the six Killing vector fields on Klein space \cite{Atanasov:2021oyu}: 
\begin{gather}
L_1= \bar{z} \partial_w + \bar{w} \partial_z, \ \ \ \ L_0=\frac{1}{2} ( z \partial_z + w \partial_w - \bar{z} \partial_{\bar{z}}- \bar{w} \partial_{\bar{w}}), \ \ \ \ L_{-1}= -z \partial_{\bar{w}}-w \partial_{\bar{z}}, \nonumber\\
\bar{L}_1= z \partial_w+\bar{w} \partial_{\bar{z}}, \ \ \ \ \ \bar{L}_0= \frac{1}{2} ( - z \partial_z + w \partial_w+ \bar{z} \partial_{\bar{z}}-\bar{w} \partial_{\bar{w}}), \ \ \ \ \ \bar{L}_{-1}=- \bar{z} \partial_{\bar{w}} - w \partial_z. 
\end{gather}

Now, using these generators, we can find the modular flow generator for celestial field theory.

Also, in $\mathbb{K}^{2,2-}$, the symmetry generators are  \cite{Atanasov:2021oyu} 
\begin{equation}
\begin{aligned}
L_1&=\frac{1}{2} e^{-i \psi - i \phi} ( \partial_\rho- i \tanh \rho \partial_\psi - i \coth \rho  \partial_\phi),\nonumber\\
L_0&=-\frac{i}{2} (\partial_\psi + \partial_\phi),\nonumber\\
L_{-1}&=\frac{1}{2} e^{i \psi + i \phi} ( -\partial_\rho- i \tanh \rho \partial_\psi - i \coth \rho \partial_\phi),\nonumber\\
\bar{L}_1&=\frac{1}{2} e^{-i \psi + i \phi} ( \partial_\rho- i \tanh \rho \partial_\psi + i \coth \rho \partial_\phi),\nonumber\\
\bar{L}_0&=-\frac{i}{2} (\partial_\psi - \partial_\phi),\nonumber\\
L_{-1}&=\frac{1}{2} e^{i \psi - i \phi} ( -\partial_\rho- i \tanh \rho \partial_\psi + i \coth \rho \partial_\phi),\nonumber\\
\end{aligned}
\end{equation}

In $\mathbb{K}^{2,2+}$, the generators are  \cite{Atanasov:2021oyu} 
\begin{equation}
\begin{aligned}
L_1&=\frac{1}{2} e^{-i \psi - i \phi} ( \partial_{\tilde{\rho}}- i \coth \tilde{\rho} \partial_\psi - i \tanh \tilde{\rho} \partial_\phi),\nonumber\\
L_0&=-\frac{i}{2} (\partial_\psi + \partial_\phi),\nonumber\\
L_{-1}&=\frac{1}{2} e^{i \psi + i \phi} ( -\partial_{\tilde{\rho}}- i \coth \tilde{\rho} \partial_\psi - i \tanh {\tilde{\rho}} \partial_\phi),\nonumber\\
\bar{L}_1&=\frac{1}{2} e^{-i \psi + i \phi} (\partial_{\tilde{\rho}}- i \coth \tilde{\rho} \partial_\psi + i \tanh \tilde{\rho} \partial_\phi),\nonumber\\
\bar{L}_0&=-\frac{i}{2} (\partial_\psi - \partial_\phi),\nonumber\\
\bar{L}_{-1}&=\frac{1}{2} e^{i \psi - i \phi} ( -\partial_{\tilde{\rho}}- i \coth \tilde{\rho} \partial_\psi + i \tanh \tilde{\rho} \partial_\phi).
\end{aligned}
\end{equation}

\begin{figure}[ht!]   
\begin{center}
\includegraphics[width=0.41\textwidth]{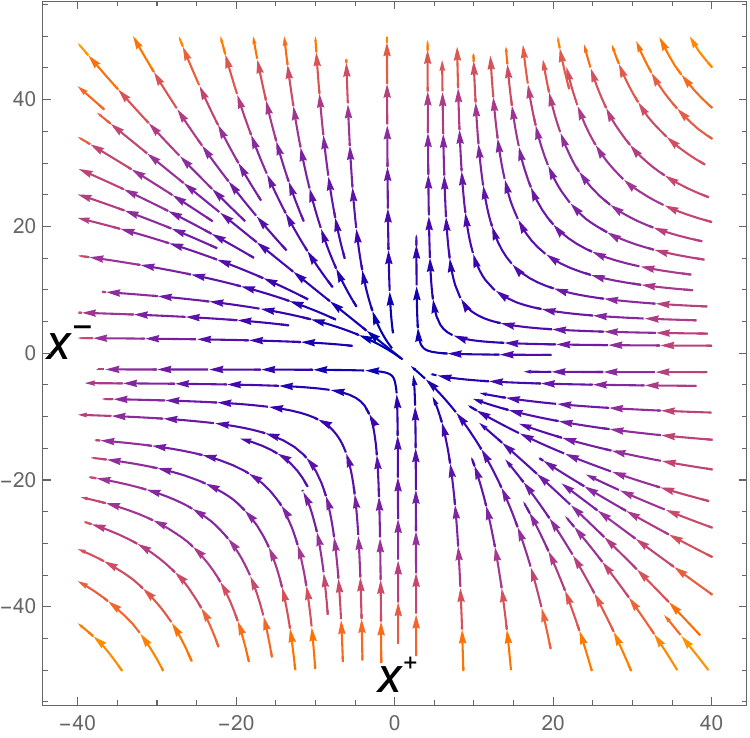} 
\includegraphics[width=0.4\textwidth]{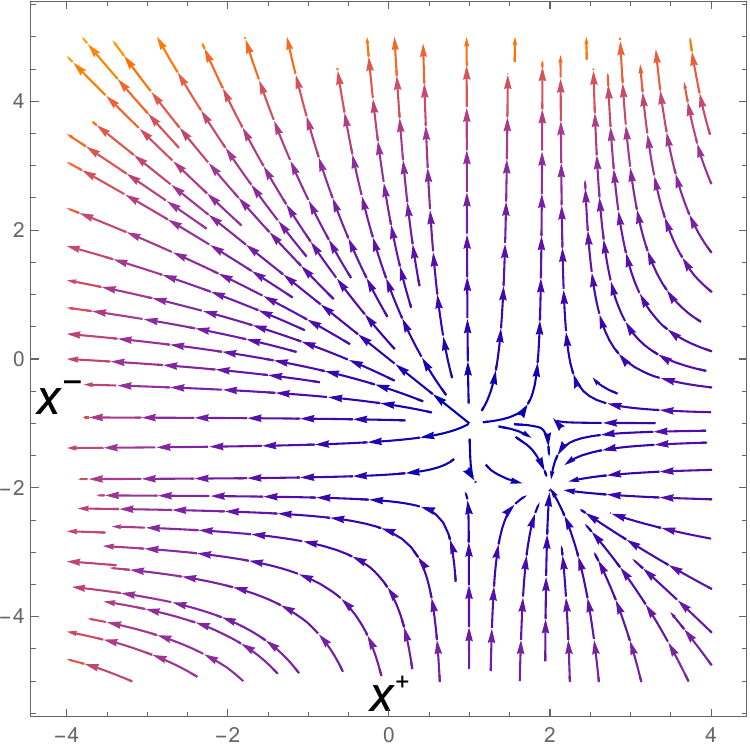} 
\caption{The modular flow of Klein CFT for an interval $\lbrack x_1^{+} , x_2^{+} \rbrack \times \lbrack x_1^{-}, x_2^{-} \rbrack$.}
\label{fig:CCFTflow}
\end{center}
\end{figure}

For the Klein CFT, or $(2,2)$ celestial torus CFT, if we take
\begin{gather}
x^{\pm} = \psi \pm \phi,
\end{gather}
the generators are
\begin{gather}
L_n= - i e^{-i n x^+} \partial_{x^+}, \nonumber\\
\bar{L}_n = - i e^{- i n (\psi - \phi) } (\partial_\psi- \partial_\phi) = -i e^{-i n x^-} \partial_{x^-}, \ \ \ n=-1,0, 1,
\end{gather}
leading to the modular flow vector field
\begin{gather}\label{eq:ccftvec}
\zeta = 2\pi \frac{(x^+ - x_1^+) (x_2^+ - x^+)}{(x_2^+ - x_1^+)} \partial_{x^+} +   2\pi \frac{(x^- - x_1^-) (x_2^- - x^-)}{(x_2^- - x_1^-)} \partial_{x^-}  ,
\end{gather}
for an interval $\lbrack x_1^{+} , x_2^{+} \rbrack \times \lbrack x_1^{-}, x_2^{-} \rbrack$ where $x^+ \in \lbrack x_1^{+} , x_2^{+} \rbrack$, $x^- \in \lbrack x_1^{-} , x_2^{-} \rbrack$. Its behavior is shown in Figure \ref{fig:CCFTflow}. Note that these generators satisfy the $SL(2, \mathbb{R})$ commutation relations.

One should note that the relation \ref{eq:ccftvec} is not periodic on the celestial torus, even though the generators constructing it are. The coordinates $(\psi, \phi)$ on the celestial torus has the periodicity of $\psi \sim \psi+2\pi$, and $\phi \sim \phi+2\pi$, which lead to $x^+ = \psi + \phi \sim x^+ + 4\pi$, and $x^- = \psi- \phi \sim x^- + 4\pi$. The components of \ref{eq:ccftvec} are parabolas in $x^\pm$ and are not periodic. So, although $\zeta$ is a valid vector field on the universal cover on the plane, it does not descend to a well-defined vector field of the torus $\mathbb{C}/ (\mathbb{Z}+ \tau \mathbb{Z})$. The resolution of this problem is that although the generators are periodic, their combination is not.  Thus, the above relation is only an approximation valid on a single patch of the universal cover. Therefore, we need another expression for the global vector field on the torus.

Note that since $L_n= -i e^{-i n x^+} \partial_{x^+}$, one can write $\partial_{x^+} = i e^{i n x^+}L_n$, and so the modular flow generator can be written as
\begin{gather}\label{eq:comb}
\xi = a_1 L_1 + a_0 L_0 + a_{-1} L_{-1} + \bar{a}_1 \bar{L}_1 + \bar{a}_0 \bar{L}_0 + \bar{a}_{-1} \bar{L}_{-1},
\end{gather}
with coefficients
\begin{gather}
a_1= \frac{2\pi i}{x_2^+ - x_1^+}, \ \ \ \ a_0 = - \frac{2\pi i (x_1^+ + x_2^+) }{x_2^+ - x_1^+}, \ \ \ \ a_{-1} = \frac{2\pi i x_1^+ x_2^+}{x_2^+ - x_1^+}, \nonumber\\
\bar{a}_1= \frac{2\pi i}{x_2^- - x_1^-}, \ \ \ \  \bar{a}_0 = - \frac{2\pi i (x_1^- + x_2^-) }{x_2^- - x_1^-}, \ \ \ \ \bar{a}_{-1} = \frac{2\pi i x_1^- x_2^-}{x_2^- - x_1^-}.
\end{gather}

Since the generators $L_n= - \frac{i}{2} e^{-i n x^+} \partial_{x^+}$, and $\bar{L}_n= - \frac{i}{2} e^{-i n x^-} \partial_{x^-}$ are proportional to $e^{-i n x^{\pm}}$, they are periodic on the torus. So for the constant coefficients $a_n$, $\bar{a}_n$, the modular flow which is a linear combination of the generators, is also manifestly periodic.  The above relation \ref{eq:comb} is a combination of periodic $SL(2, \mathbb{R})_L \times SL(2, \mathbb{R})_R $ generators given above, which reduces to the local expression \ref{eq:ccftvec} in a coordinate patch where the interval does not cross the branch cut. This is how we reconcile the manifest periodicity of the generators with the local expression for the flow.

\begin{figure}[ht!]   
\begin{center}
\includegraphics[width=0.99\textwidth]{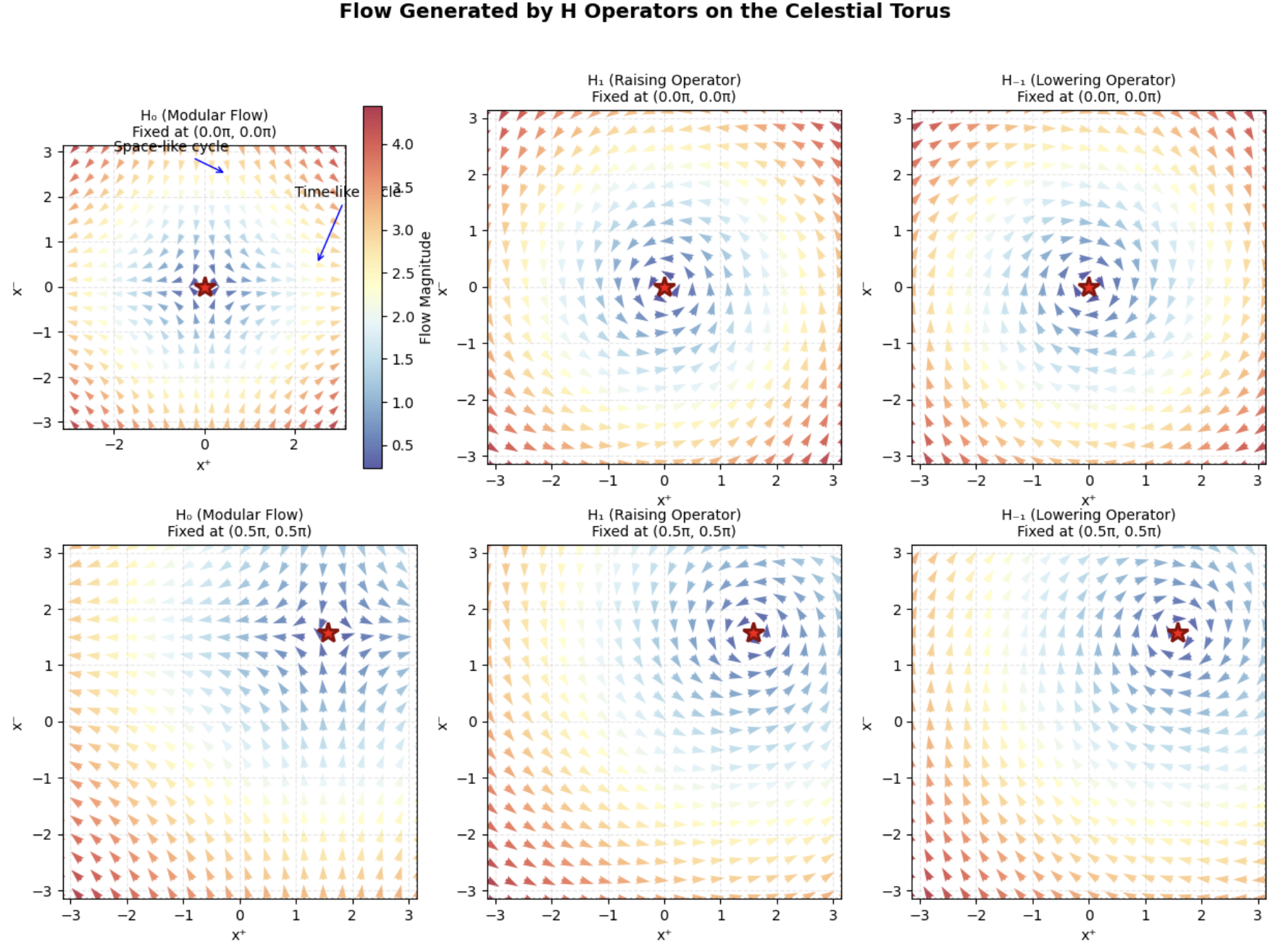} 
\caption{The flow generator by $H$ operators $H_0^{\hat{x}}$ and $H_{\pm 1}^{\hat{x}}$ on the celestial torus.}
\label{fig:Hflow}
\end{center}
\end{figure}

Note that the $H$ operators of \cite{Atanasov:2021oyu} 
 \begin{gather}
H_0^{\hat{x}}= \frac{1}{2} \left ( e^{i \hat{x}^+} L_1 - e^{- i \hat{x}^+} L_{-1}\right), \nonumber\\
H_{\pm 1}^{\hat{x}} = i L_0 \mp \frac{i}{2} \left ( e^{i \hat{x}^+} L_1 + e^{- i \hat{x}^+}L_{-1} \right),
 \end{gather}
act as modular flow generators, and their behavior is shown in Figure \ref{fig:Hflow}, which is similar to the behavior shown above.

Here, $H_0$ is the modular, or hyperbolic flow around a saddle point,  $H_1$ is the raising operator with spiral flow pattern, and $H_{-1}$ is the lowering operator with the opposite spiral flow. The red star, which is the fixed point, is where the particles emerge on the celestial torus.

The colors also show the magnitude of the modular flow, where darker colors indicate a stronger flow. In addition, $H_0$ generates the modular flow that preserves the fixed point, corresponding to equations of \cite{Atanasov:2021oyu}:
\begin{gather}
H_1^{\hat{x}} \varphi_h(X;\hat{x}) = \bar{H}_1^{\hat{x}} \varphi_h(X; \hat{x})=0,\ \ \ \ \
H_0^{\hat{x}} \varphi_h(X;\hat{x}) = \bar{H}_0^{\hat{x}} \varphi_h(X;\hat{x}) = h \varphi_h(X;\hat{x}), \nonumber\\
H_{-1}^{\hat{x}} \varphi_h(X;\hat{x}) = -2 \hat{\partial}_+ \varphi_h(X; \hat{x}),  \ \ \ \ \ \
\bar{H}_{-1}^{\hat{x}} \varphi_h(X;\hat{x}) = -2 \hat{\partial}_- \varphi_h(X; \hat{x}).
\end{gather}

Thus, having the symmetry generators of CCFT and Klein space, we can show the vector flow of the modular flow in this case.

Also, the Mellin transform, which obeys the wave equation, is
\begin{gather}
\varphi_h(X; \hat{x}) = \int_0^\infty d\omega \omega^{2h-1} e^{i \omega \hat{p}.X} = \frac{e^{-\pi i h} \Gamma(2h) }{(\hat{p} . X )^{2h}},
\end{gather}
and they have divergences for $h \in \frac{1}{2} \mathbb{Z}_-$ which is related to ``conformally soft poles".

Note that one usually considers the case where the transformations vanish at the intervals considered, and they are non-differentiable at these points; therefore, the kernel and image of adjoint action of modular Hamiltonian are decomposed.

In \cite{Czech:2019vih}, in addition to defining the modular Berry connection and modular Berry curvature using the zero modes of modular Hamiltonian, the exact forms of modular Hamiltonian and modular curvature have been calculated for the vacuum of a $\text{CFT}_2$ on a \textbf{circle}. These calculations can be performed either by using the geometry of the space of CFT intervals (the kinematic space) \cite{Czech:2017zfq},  (see also \cite{Czech:2016xec, deBoer:2016pqk}), or by using symmetries as in \cite{Czech:2019vih}. 

In the $\text{CFT}_2$ case, the set of entanglement entropies associated with intervals leads to kinematic space. Then, the kinematic space for celestial field theory can also be constructed. This space is the space of pairs of spacelike points in a $\text{CFT}_d$, and the perturbation of entanglement entropy can be seen as fields propagating in this space.

Note that the ``celestial amplitudes'' are the Mellin-transformed scattering amplitudes, and they are actually the conformal correlators of these celestial operators. So, in principle, these celestial amplitudes can be related to kinematic space and Berry phase in the setup of holographic bulk reconstruction. 

The celestial conformal field theory lives at null infinity. The main part is the celestial OPE of the two soft currents and the summation of ${SL(2,\mathbb{R})}$ descendants in the OPE. This intrinsically is actually very similar to constructing the kinematic space out of pairs of spacelike points.

\section{Discussions}

In this work, since we had the symmetry generators of CCFTs and Klein CFTs, we presented their vector flow and modular flows in figures.

Some other field theory examples that have symmetry generators different from those of Virasoro CFT could be studied. For example, Logarithmic CFT, supersymmetric field theories (SUSY QFTs), affine Kac-Moody Wess-Zumino-Witten (WZW) models, topological quantum field theories (TQFTs), non-relativistic (Lifshitz/Schrödinger) field theories, noncommutative field theories (NCFTs), higher-spin theories with $hs(\lambda)$ algebras can also be studied, and the symmetry generators and  modular flow generators can be derived.

Other very interesting field theories whose modular vector flow could be studied later are Lifshitz-type theories, Galilean conformal field theories (GCFTs), Carrollian CFT, and non-rational CFTs such as Liouville theory or specific sigma models. For Lifshitz case, we actually have a Cardy formula. 

The Cardy formula in $2d$ CFT can be written as
\begin{gather}
S_{\text{CFT}} (E) = 2\pi \sqrt{\frac{c}{6} \left(L_0 - \frac{c}{24} \right) },
\end{gather} 
which, for thermal entropy at high temperatures $T$, becomes
\begin{gather}
S_{\text{CFT}}(T) \sim \frac{\pi c}{3} T L. 
\end{gather}

For non-relativistic Lifshitz theories with dynamical critical exponent $z$ and scaling symmetry $t \to \lambda^z t$, and $x \to \lambda x$, the high-temperature entropy scales as $S(T) \sim T^{1/z}$.

So the generalized Cardy-like formula in this case is
\begin{gather}
S(T) = \frac{2\pi L}{z} \left ( \frac{E_0 z}{2\pi L} \right)^{\frac{1}{1+z} } T^{\frac{1}{z}}
\end{gather}

The holographic dual of Lifshitz field theory is Lifshitz geometry, which can be written as
\begin{gather}
ds^2 = - \frac{dt^2}{r^{2z}} + \frac{dx^2 + dr^2}{r^2},
\end{gather}
where, in $d+1$ spacetime dimensions, the black hole entropy scales as
\begin{gather}
S \sim T^{\frac{d}{z}}.
\end{gather}

The usual Cardy formula for $2d$ CFT can be written as
\begin{gather}
\rho (\Delta, \bar{\Delta})_{\text{Cardy}} \sim e^{2\pi \sqrt{\frac{c_L \Delta}{6}} } \ . \ e^{2\pi \sqrt{ \frac{c_R \bar{\Delta} }{6}} }.
\end{gather}

So, the generalized Cardy formula, which can be written as 
\begin{gather}
S(T) \sim L \ T^{1/z},  \ \ \ \ \ \ \rho(E) \sim \text{exp} ( \text{const} \ . \ E^{\frac{z}{1+z}}),
\end{gather}
would grows sub-exponentially compared to the CFT case, which grows as $\text{exp} (\sqrt{E})$.
In \cite{Loran:2010bd},  the Cardy formula was also extended and it was shown that in the saddle point approximation and up to contributions that are \textit{``exponentially suppressed"} compared to the leading Cardy's result, the density of states takes the form of $\rho(\Delta, \bar{\Delta}; c_L, c_R) = f(c_L \Delta) f(c_R \bar{\Delta} ) \rho_0(\Delta, \bar{\Delta}; c_L, c_R)$.

The modular flow and its shape dependence for excited states and real-time dynamics, as well as the gravitational dual of the modular Hamiltonian, have been studied in \cite{Faulkner:2016mzt}. The contributions to the modular Hamiltonian and modular flow from shape deformations, specifically for CCFTs and Klein CFTs, could be studied using their method.

The supersymmetric case of Einstein-Yang-Mills theory has also been recently discussed in \cite{Jiang:2021ovh}. The discussions of the Berry phase and Berry connection could be repeated for this holographic symmetry algebra. The connections between these cases and the infinite Ward identities and infinite soft currents that generate these holographic symmetry algebras could be discussed.

The OPE between a soft symmetry current and a hard operator is then summed over its $SL(2,\mathbb{R})$ descendants.  So, in general, the connections between soft theorems in scattering amplitude, Ward identities, and modular Berry phase could be studied. In the special case, the Berry phase could also be calculated for the holographic chiral algebra.

Another interesting case is the holographic chiral algebra, which can be written as \cite{Jiang:2021ovh}
\begin{gather}
\mathcal{Q}^\alpha \  . \ \mathcal{R}^{i, J}_n (z)= z^{\alpha-1} \mathcal{R}^{i, J+ \frac{1}{2}}_n (z), \ \ \ \mathcal{Q}^\alpha \ . \ \mathcal{R}^{i, J + \frac{1}{2}}_n (z) =0, \ \ \ \ J=\frac{3}{2}, \frac{1}{2}, \\
\mathcal{\tilde{Q}}^{\dot{\alpha}} \  . \ \mathcal{R}^{i, J}_n (z)= \left(i-1-n(2\dot{\alpha}-3) \right) \mathcal{R}^{i-\frac{1}{2}, J- \frac{1}{2}}_{n-\frac{3}{2}+ \dot{\alpha}} (z), \ \ \ \mathcal{\tilde{Q}}^{\dot{\alpha}} \ . \ \mathcal{R}^{i, J + \frac{1}{2}}_n (z) =0, \ \ \ \ J=2, 1. \\
\end{gather}

For this algebra, the modular zero mode and Berry connection could be discussed.

The modular zero modes change the algebra but leave the physical observables fixed. Thus, the modular connection and the non-trivial parallel transport could be calculated here.

In \cite{Guevara:2023tnf} and \cite{Guevara:2025rjh}, celestial quantum error correction has also been studied. There, they construct a toy model of CCFT from the perspective of quantum error-correcting codes. In their code, the hard states with quantized BMS soft hair form the logical subspace. This allows them to reverse errors induced by soft radiation. The connections between quantum error correction and the modular flow structure of such exotic field theories, similar to works such as \cite{Ghodrati:2020vzm, Ghodrati:2022hbb}, could be studied further, specifically for the exotic field theories we have named here.

The connection with the plane-wave formulation in hyperbolic black hole geometry of Casini, Huerta and Myers then could also be discussed. Thus, the vector fields on the circles from wave packets of these eigenstates, similar to the discussion of coadjoint orbits, and its Kirillov-Kostant symplectic form on these orbits, could be constructed.

The result for the modular Hamiltonian of a generic region in harmonic lattices has also been found in \cite{Casini:2009sr}, which could be compared with the final results of any case we obtain. In particular, the state-changing parallel transport on different bulk geometries, especially celestial bulk geometry, could be discussed.

In \cite{Javerzat:2021hxt}, numerical results from lattice models and discretization of the radial direction have been compared with the CFT calculations of the modular Hamiltonian of a sphere with massless scalar field in its ground state. They found that they match for the cases of two and three spatial dimensions and for the small values of total angular momentum. This study  can then be repeated for celestial spheres or other field theories, such as warped CFTs and BMS case.

Specifically, for the case of sphere in \cite{Javerzat:2021hxt}, they found that in its lattice model, when the mass parameter is large enough, the dominant contribution comes from the on-site and the nearest-neighbour terms, whose weight functions are straight lines. This analysis could be repeated for other cases, such as warped CFTs, which are non-local and have $SL(2,R) \times U(1)$ global symmetry algebra. Then, the dominant contribution and the form of the weight functions could be determined.
Also, one should analyze the connections between the continuum limit of lattice results for these modular Hamiltonians, the algebra, and the specific properties of each field theory. For instance, for free fermionic and bosonic systems, the ``Gaussian properties'' of the ground state lead to the match as noted in \cite{Javerzat:2021hxt}. The Gaussian-like properties could be investigated for other field theories as well. Also, the specific effects of warping factors or mass terms on the weight functions and modular Hamiltonian could be investigated.

\section*{Acknowledgments}

I would like to thank Yuan Zhong for useful discussions.

 \medskip

\bibliography{PRDBerry.bib}

@article{Czech:2019vih,
    author = "Czech, Bartlomiej and De Boer, Jan and Ge, Dongsheng and Lamprou, Lampros",
    title = "{A modular sewing kit for entanglement wedges}",
    eprint = "1903.04493",
    archivePrefix = "arXiv",
    primaryClass = "hep-th",
    doi = "10.1007/JHEP11(2019)094",
    journal = "JHEP",
    volume = "11",
    pages = "094",
    year = "2019"
}

@article{Apolo:2020bld,
    author = "Apolo, Luis and Jiang, Hongliang and Song, Wei and Zhong, Yuan",
    title = "{Swing surfaces and holographic entanglement beyond AdS/CFT}",
    eprint = "2006.10740",
    archivePrefix = "arXiv",
    primaryClass = "hep-th",
    doi = "10.1007/JHEP12(2020)064",
    journal = "JHEP",
    volume = "12",
    pages = "064",
    year = "2020"
}

@article{Ghodrati:2022lnd,
    author = "Ghodrati, Mahdis",
    title = "{Chaos, phase transitions and curvature invariants of (rotating, warped, massive) BTZ black holes}",
    eprint = "2202.05950",
    archivePrefix = "arXiv",
    primaryClass = "hep-th",
    doi = "10.1063/5.0216490",
    journal = "AIP Conf. Proc.",
    volume = "2874",
    number = "1",
    pages = "020012",
    year = "2024"
}

@article{Faulkner:2016mzt,
    author = "Faulkner, Thomas and Leigh, Robert G. and Parrikar, Onkar and Wang, Huajia",
    title = "{Modular Hamiltonians for Deformed Half-Spaces and the Averaged Null Energy Condition}",
    eprint = "1605.08072",
    archivePrefix = "arXiv",
    primaryClass = "hep-th",
    doi = "10.1007/JHEP09(2016)038",
    journal = "JHEP",
    volume = "09",
    pages = "038",
    year = "2016"
}

@article{Ghodrati:2020vzm,
    author = "Ghodrati, Mahdis",
    title = "{Entanglement wedge reconstruction and correlation measures in mixed states: Modular flows versus quantum recovery channels}",
    eprint = "2012.04386",
    archivePrefix = "arXiv",
    primaryClass = "hep-th",
    doi = "10.1103/PhysRevD.104.046004",
    journal = "Phys. Rev. D",
    volume = "104",
    number = "4",
    pages = "046004",
    year = "2021"
}

@article{Ghodrati:2022hbb,
    author = "Ghodrati, Mahdis",
    title = "{Encoded information of mixed correlations: the views from one dimension higher}",
    eprint = "2209.04548",
    archivePrefix = "arXiv",
    primaryClass = "hep-th",
    doi = "10.1007/JHEP08(2023)059",
    journal = "JHEP",
    volume = "08",
    pages = "059",
    year = "2023"
}

@article{Ghodrati:2019bzz,
    author = "Ghodrati, Mahdis",
    title = "{Complexity and emergence of warped AdS$_{3}$ space-time from chiral Liouville action}",
    eprint = "1911.03819",
    archivePrefix = "arXiv",
    primaryClass = "hep-th",
    doi = "10.1007/JHEP02(2020)052",
    journal = "JHEP",
    volume = "02",
    pages = "052",
    year = "2020"
}

@article{Ghodrati:2016vvf,
    author = "Ghodrati, M. and Hajian, K. and Setare, M. R.",
    title = "{Revisiting Conserved Charges in Higher Curvature Gravitational Theories}",
    eprint = "1606.04353",
    archivePrefix = "arXiv",
    primaryClass = "hep-th",
    reportNumber = "IPM-P-2016-020",
    doi = "10.1140/epjc/s10052-016-4550-6",
    journal = "Eur. Phys. J. C",
    volume = "76",
    number = "12",
    pages = "701",
    year = "2016"
}

@article{Ghodrati:2016ggy,
    author = "Ghodrati, Mahdis and Naseh, Ali",
    title = "{Phase transitions in Bergshoeff{\textendash}Hohm{\textendash}Townsend massive gravity}",
    eprint = "1601.04403",
    archivePrefix = "arXiv",
    primaryClass = "hep-th",
    doi = "10.1088/1361-6382/aa634f",
    journal = "Class. Quant. Grav.",
    volume = "34",
    number = "7",
    pages = "075009",
    year = "2017"
}

@article{Guevara:2025rjh,
    author = "Guevara, Alfredo and Hu, Yangrui",
    title = "{Celestial Quantum Error Correction. Part II. From qudits to celestial CFT}",
    eprint = "2412.19653",
    archivePrefix = "arXiv",
    primaryClass = "hep-th",
    doi = "10.1007/JHEP06(2025)121",
    journal = "JHEP",
    volume = "06",
    pages = "121",
    year = "2025"
}

@article{Guevara:2023tnf,
    author = "Guevara, Alfredo and Hu, Yangrui",
    title = "{Celestial quantum error correction: I. Qubits from noncommutative Klein space}",
    eprint = "2312.16298",
    archivePrefix = "arXiv",
    primaryClass = "hep-th",
    doi = "10.1088/1361-6382/adf686",
    journal = "Class. Quant. Grav.",
    volume = "42",
    number = "16",
    pages = "165006",
    year = "2025"
}

@article{Loran:2010bd,
    author = "Loran, Farhang and Sheikh-Jabbari, M. M. and Vincon, Massimiliano",
    title = "{Beyond Logarithmic Corrections to Cardy Formula}",
    eprint = "1010.3561",
    archivePrefix = "arXiv",
    primaryClass = "hep-th",
    reportNumber = "IPM-P-2010-041",
    doi = "10.1007/JHEP01(2011)110",
    journal = "JHEP",
    volume = "01",
    pages = "110",
    year = "2011"
}

@article{deBoer:2021zlm,
    author = "de Boer, Jan and Esp\'\i{}ndola, Ricardo and Najian, Bahman and Patramanis, Dimitrios and van der Heijden, Jeremy and Zukowski, Claire",
    title = "{Virasoro Entanglement Berry Phases}",
    eprint = "2111.05345",
    archivePrefix = "arXiv",
    primaryClass = "hep-th",
    month = "11",
    year = "2021"
}

@article{Jiang:2021ovh,
    author = "Jiang, Hongliang",
    title = "{Holographic Chiral Algebra: Supersymmetry, Infinite Ward Identities, and EFTs}",
    eprint = "2108.08799",
    archivePrefix = "arXiv",
    primaryClass = "hep-th",
    reportNumber = "QMUL-PH-21-36",
    month = "8",
    year = "2021"
}

@article{Donnay:2020guq,
    author = "Donnay, Laura and Pasterski, Sabrina and Puhm, Andrea",
    title = "{Asymptotic Symmetries and Celestial CFT}",
    eprint = "2005.08990",
    archivePrefix = "arXiv",
    primaryClass = "hep-th",
    reportNumber = "CPHT-RR022.042020",
    doi = "10.1007/JHEP09(2020)176",
    journal = "JHEP",
    volume = "09",
    pages = "176",
    year = "2020"
}

@article{Donnay:2021wrk,
    author = "Donnay, Laura and Ruzziconi, Romain",
    title = "{BMS Flux Algebra in Celestial Holography}",
    eprint = "2108.11969",
    archivePrefix = "arXiv",
    primaryClass = "hep-th",
    month = "8",
    year = "2021"
}

@article{Raclariu:2021zjz,
    author = "Raclariu, Ana-Maria",
    title = "{Lectures on Celestial Holography}",
    eprint = "2107.02075",
    archivePrefix = "arXiv",
    primaryClass = "hep-th",
    month = "7",
    year = "2021"
}

@article{Czech:2025jnw,
    author = "Czech, Bartlomiej and Shuai, Sirui and Wang, Yixu",
    title = "{Entropy Inequalities Constrain Holographic Erasure Correction}",
    eprint = "2502.12246",
    archivePrefix = "arXiv",
    primaryClass = "hep-th",
    doi = "10.1103/dl3c-h3hg",
    journal = "Phys. Rev. Lett.",
    volume = "135",
    number = "14",
    pages = "141603",
    year = "2025"
}

@article{Song:2016gtd,
    author = "Song, Wei and Wen, Qiang and Xu, Jianfei",
    title = "{Modifications to Holographic Entanglement Entropy in Warped CFT}",
    eprint = "1610.00727",
    archivePrefix = "arXiv",
    primaryClass = "hep-th",
    doi = "10.1007/JHEP02(2017)067",
    journal = "JHEP",
    volume = "02",
    pages = "067",
    year = "2017"
}

@article{Donnay:2022wvx,
    author = "Donnay, Laura and Fiorucci, Adrien and Herfray, Yannick and Ruzziconi, Romain",
    title = "{Bridging Carrollian and celestial holography}",
    eprint = "2212.12553",
    archivePrefix = "arXiv",
    primaryClass = "hep-th",
    doi = "10.1103/PhysRevD.107.126027",
    journal = "Phys. Rev. D",
    volume = "107",
    number = "12",
    pages = "126027",
    year = "2023"
}

@article{Donnay:2022aba,
    author = "Donnay, Laura and Fiorucci, Adrien and Herfray, Yannick and Ruzziconi, Romain",
    title = "{Carrollian Perspective on Celestial Holography}",
    eprint = "2202.04702",
    archivePrefix = "arXiv",
    primaryClass = "hep-th",
    doi = "10.1103/PhysRevLett.129.071602",
    journal = "Phys. Rev. Lett.",
    volume = "129",
    number = "7",
    pages = "071602",
    year = "2022"
}

@article{Bagchi:2022emh,
    author = "Bagchi, Arjun and Banerjee, Shamik and Basu, Rudranil and Dutta, Sudipta",
    title = "{Scattering Amplitudes: Celestial and Carrollian}",
    eprint = "2202.08438",
    archivePrefix = "arXiv",
    primaryClass = "hep-th",
    doi = "10.1103/PhysRevLett.128.241601",
    journal = "Phys. Rev. Lett.",
    volume = "128",
    number = "24",
    pages = "241601",
    year = "2022"
}

@article{Pasterski:2017kqt,
    author = "Pasterski, Sabrina and Shao, Shu-Heng",
    title = "{Conformal basis for flat space amplitudes}",
    eprint = "1705.01027",
    archivePrefix = "arXiv",
    primaryClass = "hep-th",
    doi = "10.1103/PhysRevD.96.065022",
    journal = "Phys. Rev. D",
    volume = "96",
    number = "6",
    pages = "065022",
    year = "2017"
}

@article{Apolo:2020qjm,
    author = "Apolo, Luis and Jiang, Hongliang and Song, Wei and Zhong, Yuan",
    title = "{Modular Hamiltonians in flat holography and (W)AdS/WCFT}",
    eprint = "2006.10741",
    archivePrefix = "arXiv",
    primaryClass = "hep-th",
    doi = "10.1007/JHEP09(2020)033",
    journal = "JHEP",
    volume = "09",
    pages = "033",
    year = "2020"
}

@article{deBoer:2016pqk,
    author = "de Boer, Jan and Haehl, Felix M. and Heller, Michal P. and Myers, Robert C.",
    title = "{Entanglement, holography and causal diamonds}",
    eprint = "1606.03307",
    archivePrefix = "arXiv",
    primaryClass = "hep-th",
    doi = "10.1007/JHEP08(2016)162",
    journal = "JHEP",
    volume = "08",
    pages = "162",
    year = "2016"
}

@article{Czech:2016xec,
    author = "Czech, Bartlomiej and Lamprou, Lampros and McCandlish, Samuel and Mosk, Benjamin and Sully, James",
    title = "{A Stereoscopic Look into the Bulk}",
    eprint = "1604.03110",
    archivePrefix = "arXiv",
    primaryClass = "hep-th",
    reportNumber = "SU-ITP-16-07",
    doi = "10.1007/JHEP07(2016)129",
    journal = "JHEP",
    volume = "07",
    pages = "129",
    year = "2016"
}

@article{Javerzat:2021hxt,
    author = "Javerzat, Nina and Tonni, Erik",
    title = "{On the continuum limit of the entanglement Hamiltonian of a sphere for the free massless scalar field}",
    eprint = "2111.05154",
    archivePrefix = "arXiv",
    primaryClass = "cond-mat.stat-mech",
    month = "11",
    year = "2021"
}

@article{Casini:2009sr,
    author = "Casini, H. and Huerta, M.",
    title = "{Entanglement entropy in free quantum field theory}",
    eprint = "0905.2562",
    archivePrefix = "arXiv",
    primaryClass = "hep-th",
    doi = "10.1088/1751-8113/42/50/504007",
    journal = "J. Phys. A",
    volume = "42",
    pages = "504007",
    year = "2009"
}

@article{Czech:2017zfq,
    author = "Czech, Bartlomiej and Lamprou, Lampros and Mccandlish, Samuel and Sully, James",
    title = "{Modular Berry Connection for Entangled Subregions in AdS/CFT}",
    eprint = "1712.07123",
    archivePrefix = "arXiv",
    primaryClass = "hep-th",
    doi = "10.1103/PhysRevLett.120.091601",
    journal = "Phys. Rev. Lett.",
    volume = "120",
    number = "9",
    pages = "091601",
    year = "2018"
}

@article{Atanasov:2021oyu,
    author = "Atanasov, Alexander and Ball, Adam and Melton, Walker and Raclariu, Ana-Maria and Strominger, Andrew",
    title = "{(2, 2) Scattering and the celestial torus}",
    eprint = "2101.09591",
    archivePrefix = "arXiv",
    primaryClass = "hep-th",
    doi = "10.1007/JHEP07(2021)083",
    journal = "JHEP",
    volume = "07",
    pages = "083",
    year = "2021"
}
\bibliographystyle{JHEP}
\end{document}